\documentclass[twocolumn,twocolappendix,tighten]{aastex631}
\usepackage{amssymb,amsmath,framed,cases}

\usepackage{color}

\usepackage{bm}
\expandafter\ifx\csname package@font\endcsname\relax\else
 \expandafter\expandafter
 \expandafter\usepackage
 \expandafter\expandafter
 \expandafter{\csname package@font\endcsname}
\fi
\begin{document}
\title{\large{Inferring the rate of technosignatures from 60 yr of nondetection}}

\author{Claudio Grimaldi}
\affiliation{Laboratory of Statistical Biophysics, Ecole Polytechnique F\'ed\'erale de Lausanne - EPFL, 1015 Lausanne, Switzerland}
\affiliation{Centro Studi e Ricerche Enrico Fermi, 00184 Roma, Italy}

\begin{abstract}
For about the last 60 years the search for extraterrestrial intelligence has been monitoring the sky for evidence of remotely detectable technological life beyond Earth,
with no positive results to date. While the lack of detection can be attributed to the highly incomplete sampling of the search space, 
technological emissions may be actually rare enough that we are living in a time when none cross the Earth. 
Here we explore the latter possibility and derive the likelihood of the Earth not being crossed by
signals for at least the last 60 years to infer upper bounds on their rate of emission. Under the assumption that technological emitters are distributed 
uniformly in the Milky Way and that they generate technoemissions at a constant rate, we find less than about one to five emissions generated per century 
with 95\% credible level. This implies optimistic waiting times until the next crossing event of no less than $60-1800$ years with a 
$50$\% probability. A significant fraction of highly directional signals increases the emission rates upper bounds, but without systematically changing the waiting time.
Although these probabilistic bounds are derived from a specific model and their validity depends on the model's assumptions, they are nevertheless
quite robust against weak time dependences of the emission rate or nonuniform spatial distributions of the emitters.
Our results provide therefore a benchmark for assessing the lack of detection and may serve as a basis to form optimal strategies for the search 
for extraterrestrial intelligence.
\end{abstract}

\section{Introduction}
\label{SecIntro}
Searching for a needle in a ``cosmic haystack" is a catchy metaphor that vividly illustrates the difficulties encountered by the search for extraterrestrial intelligence (SETI)
due to the vastness of the parameter space to be searched \citep{Tarter2010,Wright2018}. 
Hypothetical technological species might indeed manifest themselves, either intentionally or not, 
through electromagnetic emissions reaching our planet from unknown locations in space and with wavelength, radiated power, duration and other transmission 
characteristics of which we have no prior knowledge \citep{Forgan2019,Lingam2021}.
To get an idea of the vastness of the search space, \citet{Tarter2010} compared the fraction of parameter space explored during the first 50 years of SETI as
equivalent to $1.6$ cups of water from Earth's oceans. After a decade and many other surveys, \citet{Wright2018} updated this estimate 
by replacing the $1.6$ cups of water with a small swimming pool; still a tiny fraction of Earth's oceans. 

Despite over 60 years of activity, it is thus not surprising that the search for extraterrestrial intelligence, or more properly the search 
for remotely detectable manifestations of technology (also known as technosignatures), has so far ended up empty-handed. 
The strategy behind SETI's ongoing efforts, then, is to continually improve the sampled search space through increasingly comprehensive 
surveys, such as the Breakthrough Listen initiative \citep{Worden2017}, or to consider technosignatures more exotic than radio or optical \citep{Sellers2022}
with the hope of eventually finding the 
long-sought needle in the cosmic haystack, or at least placing even tighter upper limits on its existence \citep{Enriquez2017,Grimaldi2018,Price2020,Wlodarczyk2020,Gajjar2021,Gajjar2022,Suazo2022}.

Although the elusiveness of extraterrestrial technosignatures might be justified by the aforementioned immense search space to be explored,
 it is however also consistent with the possibility that there are actually no technosignatures to be detected.
This does not necessarily mean that technological exo-civilizations or their emitting artifacts are extremely rare or nonexistent \citep{Tipler1980,Ward2000}, 
but, less categorically, that we are looking for them during a time when our planet is in a region of the galaxy devoid of technoemissions, even if other regions are illuminated by them. This could be, for example, the case of extraterrestrial emitters that have generated electromagnetic radiations propagating in all directions at the 
speed of light, but that have not yet reached our planet, or that have ceased radiating in a past sufficiently distant that their signals
have already overcome the Earth and continue to move away from it. If there is a non-zero emission rate, then, this scenario implies that 
while the signals that are moving away from our planet will remain forever invisible to us, others are heading our way
and will be potentially detectable in the future when they eventually cross the Earth.

Here, we investigate the consequences of assuming that the Earth has not been crossed by any technosignal at least since humanity 
began to actively search for them. Although sporadic searches for (radio) signals predated the first modern SETI experiment, conducted 
in 1960 \citep{Drake1961}, we take a fiducial period of $60$ years of non-detection as our working hypothesis. 
As shown in the following, this strategy allows us to place upper limits
on the rate of technoemissions and to infer probabilistic waiting times until the next crossing event, without recurring to additional hypotheses about 
emission longevities or other emission characteristics.

\begin{figure}[t]
\includegraphics[width=8.5 cm]{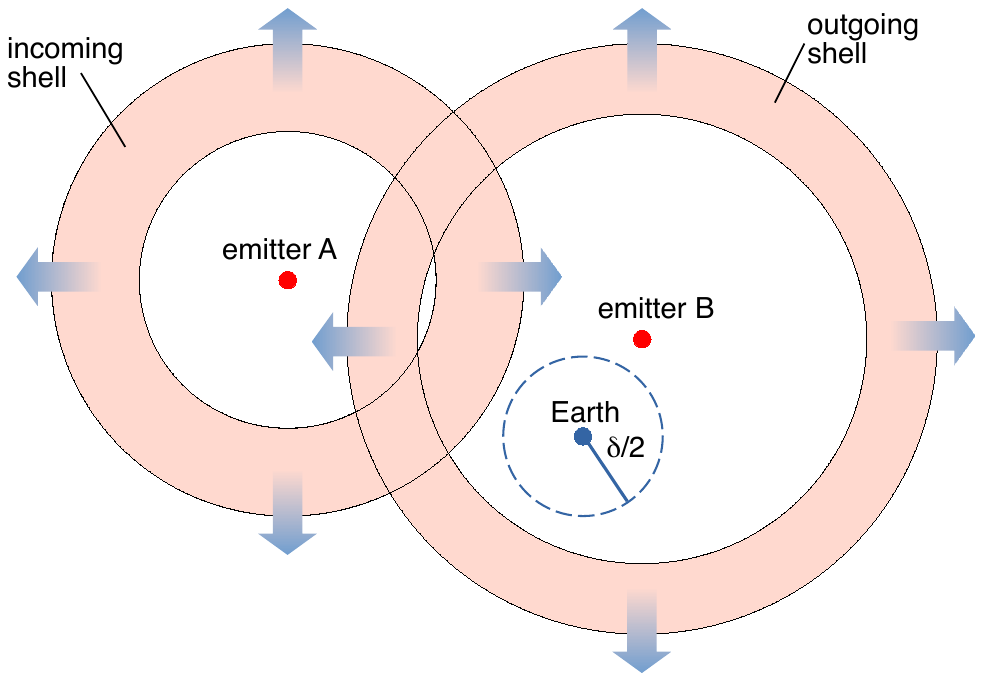}\\
\caption{Spherical shell model of isotropic technoemissions. The two annular regions are a two-dimensional representation of the space covered by the isotropic 
radiations originating from emitters A and B. The thicknesses of the annuli are proportional to the emission longevities, whereas the outer radii are proportional to
the time elapsed since the beginning of the emission processes. The arrows indicate the direction 
of propagation at the speed of light $c$ of the outer and inner edges of the annuli. The dashed circle represents a test sphere of radius $\delta/2$ and with 
center at Earth's position. The time interval between successive overlap events between the shells and the test sphere is greater than $\tau=\delta/c$.}
\label{fig1}
\end{figure}

\section{The model}
\label{sec:model}
In what follows, we refer to an ``emitter" as any extraterrestrial source of artificial electromagnetic emissions, 
regardless of whether the source is an actively transmitting technological civilization, a robotic transmitter, or the byproduct of some 
technological activity. We assume that such emitters are independently and identically distributed in the Milky Way Galaxy with probability 
distribution function (PDF) $\rho_E(\mathbf{r})$, where $\mathbf{r}$ is the emitter position relative to the galactic center. 
Since here we are interested in a scenario where the Earth is in a region of space devoid of technoemissions, hereafter referred to as the void space, 
we do not specify characteristics such as wavelength, intensity, duty-cycle, etc., and only assume
that the emissions are generated at a constant (i.e., time-independent) rate per unit 
volume $\Gamma\rho(\mathbf{r})$, where $\Gamma$ is the emission birthrate in the entire Galaxy. We defer the discussion about the validity 
of this approximation to the end of this Section.

Let us first treat the case of isotropic technoemissions, since the case of directional, anisotropic emissions can be derived directly from the isotropic one, 
as shown in Sec.~\ref{sec:anis}.

Examples of isotropic technoemissions are the infrared glow generated by hypothetical mega-structures, such as the
Dyson spheres \citep{Dyson1960}, the radio or optical emissions from beacons sweeping the entire galaxy, or leaked electromagnetic radiations
produced by technological activities. In principle, this list could also include remotely detectable industrial pollution in the atmosphere of exoplanets \citep{Lin2014,Kopparapu2021}, although searches of this kind have not yet been carried out.

We model the region of space filled by an isotropic emission process lasting a time interval $L$ with a spherical shell centered at $\mathbf{r}$ 
and having outer radius $ct$ and thickness $cL$, where $c$ is the speed of light and $t$ is the time elapsed since the beginning of the 
emission process \citep{Smith2009,Grimaldi2018}.
As mentioned above, the emissions are generated at a constant rate, so at any given time the galaxy is filled with a certain number of spherical shells 
with uniformly distributed outer radii. We make the further reasonable assumption that the durations of the emission processes 
(or, equivalently, the thicknesses of the spherical shells) are independently and identically distributed random variables with PDFs 
given by $\rho_L(L)$.

We now focus on the aforementioned scenario where none of the emissions present in the Galaxy crosses the Earth.
As shown in Figure~\ref{fig1}, we can identify two types of shells for this to happen: the incoming and the outgoing shells. 
The shells of the first type have an outer radius that is larger than the distance of the Earth from their point of origin, as the shell generated by the emitter A in Fig. \ref{fig1}.
Since the outer shell radii are expanding at the speed of light, the incoming shells will reach the Earth 
at some time in the future. The second type of shells, the outgoing shells, are such that the Earth is located within their ``hole",
as in the case of the shell generated by emitter B in Fig. \ref{fig1}. In this case, the outgoing shells are steadily moving away from our planet and  
have overlapped the Earth at some time in the past. 

\begin{figure*}[t]
\begin{center}
\includegraphics[width=18 cm]{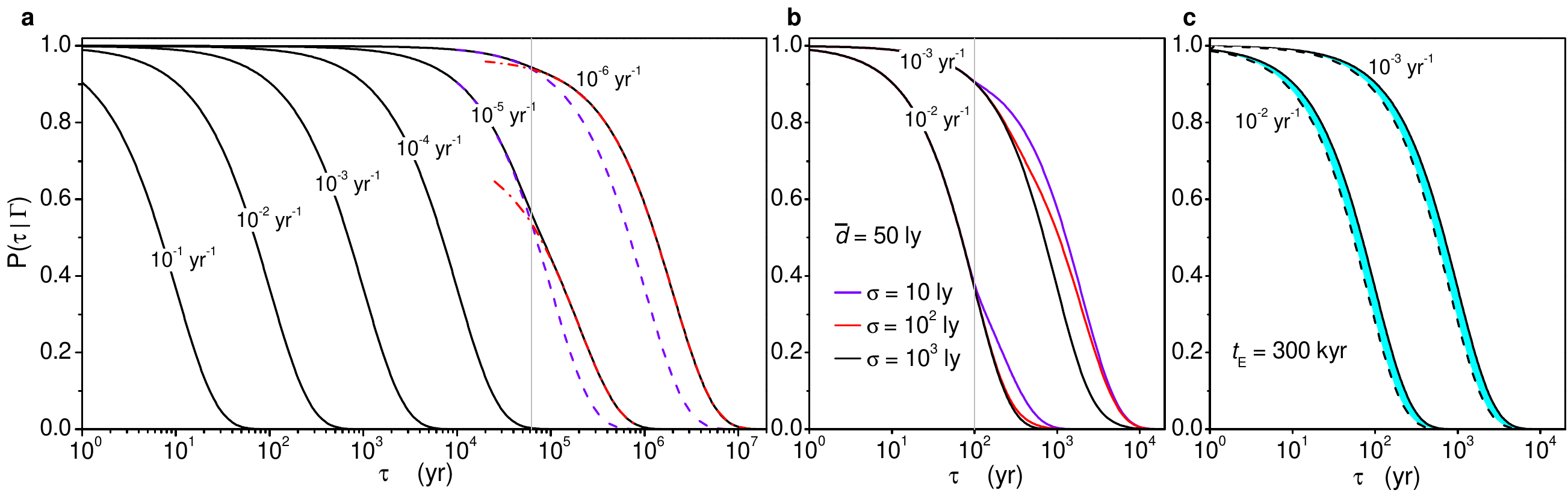}
\caption{Likelihood $P(\tau\vert\Gamma)$ of Earth-technosignal noncrossing time being greater than $\tau$. a: black solid lines represent the conditional probability
$P(\tau\vert\Gamma)$ calculated numerically from Equation~\eqref{prob3} for different values of the emission birth rate $\Gamma$ and assuming that the emitters are
distributed uniformly over the thin disk of the Milky Way. For $\tau$ smaller than and greater than about $2\bar{d}/c\simeq 62$ kyr (vertical gray line), 
$P(\tau\vert\Gamma)$ is well approximated by $\exp(-\Gamma\tau)$ (blue dashed lines) and $ \exp[-\Gamma(\tau/2+\bar{d}/c)]$ (red dot-dashed lines), respectively. 
(b) $P(\tau\vert\Gamma)$ calculated for emitters that are normally distributed around a distance of $50$ ly from Earth for different values of the standard
deviation $\sigma$. The vertical gray line denotes $2\bar{d}/c=100$ yr. c: The solid lines are the likelihoods in the stationary limit for $\Gamma= 10^{-3}$ and
$10^{-2}$ yr$^{-1}$ (as in panel a), whereas the dashed lines are the corresponding likelihoods computed in the non-stationary case for $t_E=300$ kyr 
and $\bar{L}=100$ kyr. The solid and dashed lines encompass the results for all values of $\bar{L}< 100$ kyr and $t_E>300$ kyr.}
\label{fig2}
\end{center}
\end{figure*}

To estimate the typical time interval between two crossing events, and therefore the typical time during which the Earth is located in a void space, we resort
to a method similar to that used in soft matter to characterize the void or pore space in porous media or, more generally, in two-component materials \citep{Torquato2002}.
Namely, we treat the Earth as if it were the center of a test sphere of diameter $\delta\geq 0$ and consider the probability 
that none of the spherical shells overlaps the test sphere: 
\begin{equation}
\label{prob1}
P(\delta)=e^{-\eta(\delta)},
\end{equation}
where $\eta(\delta)$ denotes the average number of shells overlapping the test sphere. Since $P(0)$ is the probability of the 
Earth being in the void space, $P(\delta)/P(0)$ gives the expected fraction of the void space available to the test sphere, also known
as the cumulative pore-size distribution function \citep{Torquato2002}.

Now, $P(\delta)/P(0)$ is also equivalent to the probability that
the outer radius of the nearest incoming shell and the inner radius of the nearest outgoing shell are each at a distance not smaller than $\delta/2$ 
from the Earth and, consequently, for a given emission rate the time interval between successive overlaps has a probability 
\begin{equation}
\label{prob2}
P(\tau\vert\Gamma)\equiv P(c\tau)/P(0)=e^{\eta(0)-\eta(c\tau)}
\end{equation}
of being greater than $\tau=\delta/c$.

We calculate $\eta(c\tau)$ as described in the Appendix \ref{AppA} to find:
\begin{align}
\label{etaa}
\eta(c\tau)=\Gamma\bigg[\tau+\bar{L}-\frac{1}{2}\int\!d\mathbf{r}\,&\rho_E(\mathbf{r})\theta(\tau-2\vert\mathbf{r}-\mathbf{r}_o\vert/c)\nonumber \\
&\times(\tau-2\vert\mathbf{r}-\mathbf{r}_o\vert/c)\bigg],
\end{align}
where $\theta$ is the unit step function, $\mathbf{r}_o$ is the vector position of the Earth, and $\bar{L}=\int\! dL\,\rho_L(L)L$ is the average longevity of the emission 
processes, a key factor in determining the probability of contact \citep{Lares2020,Kipping2020,Balbi2021}. 

A first critical result is that, since $\eta(0)=\Gamma\bar{L}$, the average longevity cancels out in $P(\tau\vert\Gamma)$. 
This is beneficial for the analysis that follows because $\bar{L}$ is an utterly 
unknown parameter whose value has been the subject of much speculation since the early days of SETI \citep{Shklovskii1965,Gott1993,Wright2022}.

\begin{figure*}[t]
\begin{center}
\includegraphics[width=18 cm]{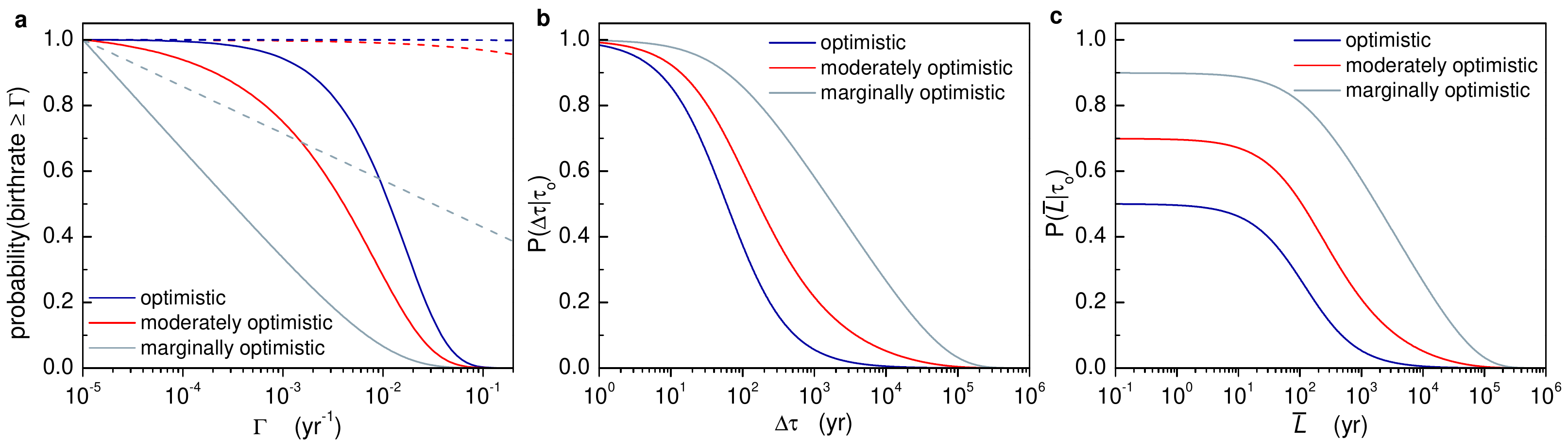}
\caption{Posterior probabilities from the $60$-year long absence of technosignals at Earth. 
a, Posterior probability of the emission rate being greater than $\Gamma$ (solid lines)
inferred from three different priors (dashed lines): optimistic (PDF uniform in $\Gamma$), moderately optimistic (PDF uniform in $\sqrt{\Gamma}$),
and marginally optimistic (PDF uniform in $\log\Gamma$). b, Posterior probability of the next crossing event occurring not sooner than $\Delta\tau$ calculated
from Equation \eqref{Ptauo} for the three optimistic cases. c, Posterior probability of the average emission longevity $\bar{L}$ calculated from Equation \eqref{bayes3}.}
\label{fig3}
\end{center}
\end{figure*}

After eliminating $\bar{L}$, two unknowns are left in $P(\tau\vert\Gamma)$: the emission birth rate, $\Gamma$, and the spatial distribution 
of the emitters, encoded by $\rho_E(\mathbf{r})$ in Equation \eqref{etaa}. In modeling the latter, we assume that the emitters do not occupy a special region 
of the galaxy and adopt for $\rho_E(\mathbf{r})$ an axisymmetric PDF that reproduces the distribution of stars in the thin disk of the Milky Way 
(see the Appendix \ref{AppA3} for more details). 

Figure \ref{fig2}(a) shows $P(\tau\vert\Gamma)$ as a function of $\tau$ for several values of the emission rate $\Gamma$ calculated numerically from 
Equations \eqref{prob2} and \eqref{etaa} using $r_o=27$ kly. As shown in the figure, $P(\tau\vert\Gamma)$ closely follows
\begin{subnumcases}{\label{etab} P(\tau\vert\Gamma)\simeq}
e^{-\Gamma\tau} & \textrm{for  $\tau\leq 2\bar{d}/c$}, \label{etab1}\\
e^{-\Gamma(\tau/2+\bar{d}/c)} & \textrm{for $\tau> 2\bar{d}/c$},\label{etab2}
\end{subnumcases}
where $\bar{d}=\int\!d\mathbf{r}\rho_E(\mathbf{r})\vert\mathbf{r}-\mathbf{r}_o\vert$ is the average distance of an emitter from the Earth, which is about 31.08 kly
for the emitter distribution here considered. Therefore, as long as $\tau\lesssim 6.2\times 10^4$ yr (vertical gray line in Fig.\ref{fig2}(a)) $P(\tau\vert\Gamma)$ 
essentially coincides with the probability of the waiting time between events of a Poisson point process with rate parameter $\Gamma$. 

Our minimal model of uniform distribution of emitters in the Galaxy can be generalized to consider other density profiles, such as
the annular galactic habitable zone of \citet{Lineweaver2004}, which gives essentially the same results of Fig.~\ref{fig2}(a), or much less uniform
ones, such as those describing emitters that are clustered in more or less localized regions of the Galaxy. For the sake of illustration, Fig.~\ref{fig2}(b)
shows the conditional probability $P(\tau\vert\Gamma)$ calculated by adopting for $\rho_E(\mathbf{r})$ a Gaussian of dispersion $\sigma$  and 
mean $\mathbf{r}_E$, such that $\vert \mathbf{r}_o-\mathbf{r}_E\vert= 50$ ly. When the emitter distribution localizes more tightly around $\mathbf{r}_E$
(small $\sigma$), the conditional probability approaches the piecewise functional form of Eq.~\eqref{etab}.
 
At this stage, a few remarks should be made about the assumption of a stationary birthrate of technoemissions. This assumption has often been
questioned on the basis that the habitability of the Galaxy is, itself, a function of time \citep{Lineweaver2004}, so that it is reasonable that $\Gamma$ also 
varies with $t$ \citep{Cirkovic2004,Balbi2021}, albeit over an unknown timescale $t_E$.
Here, we do not delve into speculation about what the temporal dependence of $\Gamma(t)$ might be, but rather estimate the timescale $t_E$
such that the temporal dependence of the birthrate can be neglected. To this end, we expand $\Gamma(t)$ up to the first order in $t$, 
$\Gamma(t)\simeq\Gamma(1+t/t_E)$ \citep{Balbi2022}, and calculate the resulting $P(\tau\vert\Gamma)$ as outlined in the Appendix \ref{AppA2}.
We find that $P(\tau\vert\Gamma)$ reduces to the stationary limit as long as both $\bar{d}/c$ and the average longevity $\bar{L}$ are
much smaller than $t_E$. For a uniform distribution of the emitters ($\bar{d}/c\simeq 31$ kyr) a stationary birthrate is thus a good approximation when $t_E$ is 
greater and $\bar{L}$ is smaller than about $10\bar{d}/c\simeq 300$ kyr, as shown by the numerical results plotted in Fig. \ref{fig2}(c).

\section{Results}
\label{sec:results}

We now turn to the implications of assuming that the fruitless efforts during the $\sim 60$-year history of SETI are actually due to the absence 
of Earth-shell overlaps for at least $\tau_o=60$ yr, rather than to a highly incomplete sampling of the search space. Keeping in mind the caveats 
in the previous section, in the following we consider the emitters to be generated at a constant rate and uniformly distributed over the Milky Way.

\subsection{Inferred emission rates}
\label{sec:emission}
We start by inferring the posterior probability distribution of $\Gamma$ using Bayes' theorem:
\begin{equation}
\label{bayes1}
p(\Gamma\vert\tau_o)=\frac{P(\tau_o\vert\Gamma)p(\Gamma)}{\int\!d\Gamma P(\tau_o\vert\Gamma)p(\Gamma)},
\end{equation}
where $p(\Gamma)$ is the prior PDF of $\Gamma$ representing some initial hypothesis about the emission birth rate and  
$P(\tau_o\vert\Gamma)=P(c\tau_o)/P(0)$ is the likelihood that the time interval between overlaps is greater than $\tau_0$, given $\Gamma$. 
We use Equation~\eqref{etab1} for $P(\tau_o\vert\Gamma)$, which is justified by the small value of $\tau_o$:
\begin{equation}
\label{bayes2}
p(\Gamma\vert\tau_o)=\frac{e^{-\Gamma\tau_o}p(\Gamma)}{\int\!d\Gamma e^{-\Gamma\tau_o}p(\Gamma)},
\end{equation}
which shows that values of $\Gamma$ much greater than $1/\tau_o\sim 0.02$ yr$^{-1}$ are strongly disfavored. This implies that it is unlikely 
that far more than two shells per century are emitted from the Milky Way and that, consequently, an \textit{a priori} optimistic view asserting a high rate 
of emissions must be significantly reconsidered.

To place more quantitative upper bounds on $\Gamma$, we adopt three different functional forms of the prior that reflect distinct shades of optimism towards the 
possible emission rate: a prior PDF uniform in $\Gamma$, a prior uniform in $\sqrt{\Gamma}$, 
and a prior uniform in the logarithm of $\Gamma$. All three priors are defined in the interval 
$\Gamma_\textrm{min}=10^{-5}$ yr$^{-1}$ to $\Gamma_\textrm{max}= 10^2$ yr$^{-1}$ (and $0$ otherwise).
The uniform in $\Gamma$ and uniform in $\sqrt{\Gamma}$ priors represent, respectively, an optimistic and a moderately optimistic belief 
about the emission birth rate, 
as they assert, for example, that $\Gamma < 10^{-2}$ yr$^{-1}$ is respectively $100$ times and $10$ times less likely than $\Gamma < 1$ yr$^{-1}$.
Conversely, the log-uniform prior is in principle uninformative, as it implies almost complete ignorance of even the scale of $\Gamma$ \citep{Spiegel2012}. 
However, the lower limit of $\Gamma$ set at $10^{-5}$ yr$^{-1}$ assumes the presence at any time of at least $\sim 1$ spherical shell within the galaxy \citep{Grimaldi2021}, making even the log-uniform prior at least marginally optimistic.

Figure ~\ref{fig3}(a) shows the posterior probability of the emission rate being larger than $\Gamma$, $P(\Gamma\vert\tau_o)$, calculated by integrating 
Equation~\eqref{bayes2}  from $\Gamma$ to $\Gamma_\textrm{max}$. Depending on the degree of optimism transpiring from the priors, 
the assumption that no technoemissions have crossed the Earth during (at least) 
the entire history of SETI implies that $\Gamma$ is less than about $0.05$ yr$^{-1}$ (optimistic), $0.03$ yr$^{-1}$ (moderately optimistic) 
and $0.01$ yr$^{-1}$ (marginally optimistic) with a credible level of $95$\%. Overall, this translates into an upper bound of about one to five emissions per 
century generated throughout the galaxy, roughly corresponding  to the inferred rate of supernovae in the Milky Way \citep{Rozwadowska2021}.

This estimate does not change much even in the extreme case of emitters strongly localized at only $10$ ly from Earth, in which case we infer using 
Equation \eqref{etab2} an upper bound on $\Gamma$ of about two to seven emissions per century.

\subsection{Waiting time}
\label{sec:waiting}
Having established that we can infer information on $\Gamma$ directly from the $60$-year-long absence of Earth-shell overlaps, we now show that this can 
be used to inform us about the waiting time $\Delta\tau$ until the next overlap event. To this end, we take
 the conditional probability of no overlap during a time interval of at least $\tau_o+\Delta\tau$ years, given that no overlap has persisted for at least
 $\tau_o$ years: $P(\tau_o+\Delta\tau\vert\Gamma)/P(\tau_o\vert\Gamma)=\exp(-\Gamma\Delta\tau)$. Marginalization over $p(\Gamma\vert\tau_o)$ yields:
\begin{equation}
\label{Ptauo}
P(\Delta\tau\vert\tau_o)=\int\!d\Gamma e^{-\Gamma\Delta\tau}p(\Gamma\vert\tau_o)=\frac{\int\!d\Gamma e^{-\Gamma(\Delta\tau+\tau_o)}p(\Gamma)}
{\int\!d\Gamma e^{-\Gamma\tau_o}p(\Gamma)},
\end{equation}
from which we derive that the median of $P(\Delta\tau\vert\tau_o)$ is about $60$ yr (optimistic), $170$ yr (moderately optimistic) 
and $1800$ yr (marginally optimistic). Even in the most optimistic case there is a decent $20$\% probability that the next crossing event will
occur not sooner than $240$ yr (Figure~\ref{fig3}(b)), whereas we can be confident that in the least optimistic scenario the waiting time does not exceed
about $10^5$ yr ($95$\% credible level). This is due to our choice of setting $\Gamma_\textrm{min}=10^{-5}$ yr$^{-1}$ for the minimum emission rate,
which prevents the log-uniform prior to diverge as $\Gamma\rightarrow 0$. Smaller values of $\Gamma_\textrm{min}$ (hence more pessimistic log-uniform priors)
would result in longer waiting times than those inferred from the marginally optimistic case of Figure~\ref{fig3}(b).

A word of caution is in order regarding the fallacy of interpreting $\Delta\tau$ as the expected waiting time until a possible future detection. 
In fact, $P(\Delta\tau\vert\tau_o)$ gives the temporal scale associated to the non-overlap with technoemissions, regardless of whether detectors on
Earth actively search for them. Because of the aforementioned vastness of the search space, perspectives on the actual detection of technosignatures, 
therefore, pertain to time scales that are necessarily larger than those predicted by $P(\Delta\tau\vert\tau_o)$, 

\begin{figure*}[t]
\begin{center}
\includegraphics[width=18 cm]{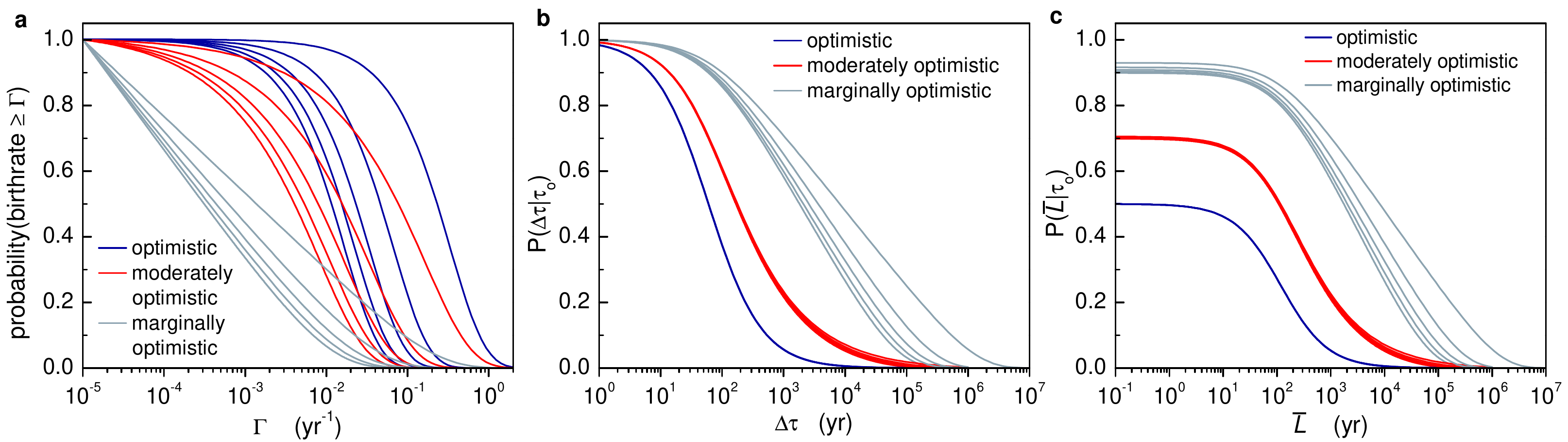}
\caption{Effects of technoemission anisotropy on the posterior probabilities. 
a, Posterior probability of the emission rate being greater than $\Gamma$ for different fractions $q$ of anisotropic technoemissions modelled by randomly 
oriented narrow beams with aperture of $2$ arcmin ($\alpha\simeq 6\times 10^{-4}$ rad). For each prior considered $q=0$, $0.25$, $0.5$, $0.75$, and 
$0.95$ (from left to right). b, Corresponding posterior probability of the next crossing event occurring not sooner than $\Delta\tau$. 
c, Posterior probability of the average emission longevity $\bar{L}$. }
\label{fig4}
\end{center}
\end{figure*}

\subsection{Inferred longevities}
\label{sec:long}
So far we have assumed that no spherical shell has intersected the Earth for at least $60$ years. But how likely is this scenario 
in light of the emission rates inferred in Sec.\ref{sec:emission}? To find it out we consider the probability of the test sphere not crossing any 
shell signal, given in Equation~\eqref{prob1}.
Neglecting again the integral term in Equation~\eqref{etaa} we obtain $P(c\tau_o)=\exp[-\Gamma(\tau_o+\bar{L})]$,
where the exponential drop with $\bar{L}$ reflects the narrowing of the voids as the signal longevity increases. 
Marginalization over the posterior PDF of $\Gamma$ gives
\begin{equation}
\label{bayes3}
P(\bar{L}\vert\tau_o)=\int\!d\Gamma e^{-\Gamma(\tau_o+\bar{L})}p(\Gamma\vert\tau_o)
=\frac{\int\!d\Gamma e^{-\Gamma(2\tau_o+\bar{L})}p(\Gamma)}
{\int\!d\Gamma e^{-\Gamma\tau_o}p(\Gamma)},
\end{equation}
which is plotted in Figure \ref{fig3}(c) for the three different priors considered. We found that for $\bar{L}\lesssim 2\tau_o=120$ yr the non-overlap probability 
$P(\bar{L}\vert\tau_o)$ is over $25$\% (optimistic), $48$\% (moderately optimistic)
and $80$\% (marginally optimistic). Interestingly, technoemissions need not be short-lived to allow for a non-overlap period of $>60$ years, as their longevity 
can reach $1,100$ and $16,600$ yr with an appreciable $20$\% probability for the moderately and marginally optimistic cases, respectively (Figure~\ref{fig3}(c)).
However, as a consequence of assuming $\Gamma_\textrm{min}=10^5 $yr$^{-1}$, even the least optimist scenario rules out average longevities greater 
than about $10^5$ yr.

\subsection{Anisotropic emissions}
\label{sec:anis}
Now, we elaborate on the possibility that a fraction $q$ of technoemissions is given by more or less long-lived directional signals, such as collimated radio beams or
optical and infrared laser signals \citep{Townes1983,Tellis2015}. In this case, the total emission rate can be written as $\Gamma=\Gamma_\textrm{iso}+\Gamma_\textrm{ani}$,
where $\Gamma_\textrm{iso}$ and $\Gamma_\textrm{ani}$ are respectively the rates of isotropic and anisotropic technoemissions with corresponding average longevities
given by $\bar{L}_\textrm{iso}$ and $\bar{L}_\textrm{ani}$, and $q=\Gamma_\textrm{ani}/\Gamma$. Since the space filled by the radiation of a directional signal
is smaller than that occupied by an isotropic emission of similar longevity, we expect an increased average size of the void regions as $q\neq 0$ .
To see this, we model the anisotropic emissions by narrow conical beams of angular aperture $\alpha\ll 2\pi$ and beam axis orientations distributed uniformly 
over the unit sphere. This model is in principle suitable for describing directional signals aimed at targets other than Earth, but that could 
accidentally illuminate it. As shown in Appendix \ref{AppB}, Equation \eqref{etab} still gives the probability of the time interval between overlaps 
being greater than $\tau$, provided that we adopt for $\Gamma$ the effective rate $\Gamma^*=\Gamma\chi$, where $\chi=[(1-q)+q\alpha^2/16]\leq 1$ 
accounts for the enlarged space available to the test sphere \citep{Grimaldi2021}.

The use of the likelihood function $P(\tau_o\vert\Gamma^*)=\exp(-\Gamma^*\tau_o)$ allows us to compute the posterior probabilities along the same lines
described above for the isotropic case. As summarized in Figure \ref{fig4}(a), the posterior probability of $\Gamma$ increases as $q>0$
(with $\alpha$ held fixed at 2 arcmin $\simeq 6\times10^{-4}$ rad) for the three optimistic scenarios considered. For example,
assuming that half of the emissions are generated by randomly oriented narrow beams ($q= 50$ \%), the inferred total emission 
rate turns out to be less than $0.02$--$0.1$ yr$^{-1}$ (from the least to the most optimistic scenarios) with a credible level of $95$\%, 
thus doubling the probabilistic upper bounds found for totally isotropic technoemissions. 

The increase in the posterior probability of $\Gamma$, however, has virtually no effect on the posterior probabilities of $\Delta\tau$ and $\bar{L}$ 
in the optimistic and moderately optimistic scenarios (Figures \ref{fig4}(b) and \ref{fig4}(c)), because such an increase
is almost completely compensated by the decrease of the anisotropy factor $\chi$. The compensation becomes complete if we take $\Gamma_\textrm{min}=0$ yr$^{-1}$
and $\Gamma_\textrm{max}=\infty$, for which we find $P(\Delta\tau\vert\tau_o)=\tau_o/(\Delta\tau+\tau_o)$ 
and $P(\bar{L}\vert\tau_o)=\tau_o/(\bar{L}+2\tau_o)$ in the optimistic case and
$P(\Delta\tau\vert\tau_o)=\sqrt{\tau_o/(\Delta\tau+\tau_o)}$ and $P(\bar{L}\vert\tau_o)=\sqrt{\tau_o/(\bar{L}+2\tau_o)}$ in the moderately optimistic case.
On the contrary, the divergence of the log-uniform prior PDF for $\Gamma\rightarrow 0$ makes the posteriors of $\Delta\tau$ and $\bar{L}$ still dependent of
the anisotropy factor $\chi$ (Figures \ref{fig4}(b) and (c)).

\section{Discussion and conclusions}
\label{sec:concl}
We have presented the results of the hypothesis that our planet has not been crossed 
by extraterrestrial technological emissions for at least $60$ years, corresponding to the period when SETI has been actively (albeit intermittently) 
searching for technosignatures. Although the lack of detection to date can be justified by the highly incomplete sampling of the SETI search space,
our working hypothesis is consistent with the available data and represents a much less worst-case scenario for SETI science than the 
claims of extreme rarity or even total absence of technological species other than ours to explain why they have not been detected so far.

Borrowing a formalism pertaining to soft matter physics and using standard Bayesian methods, we inferred upper bounds on the 
technoemission rate $\Gamma$ and corresponding lower bounds on the waiting time until the next crossing event that are remarkably independent of the
signal longevity. We have shown that if the lack of detection for the past $60$ years happens to be due to our planet being in a 
region devoid of technosignals, then it follows that SETI will likely find none 
for the coming several decades (if not centuries or even millennia for the least optimistic case), even if it were to search "all-sky, all-the-time". 

This conclusion rests on a few assumptions we made regarding the emission rate of technosignals and the spatial distribution of the emitters,
which we will now comment on. We start by noting that relaxing the hypothesis that the emissions are generated at a constant rate would 
make our central quantity, the likelihood function $P(\tau\vert\Gamma)$,
dependent on the emission longevity $L$. This implies that additional assumptions about the $L$ distribution are needed to infer the waiting time 
until the next crossing event. However, we have shown that as long as $\Gamma(t)$ varies over time scales $t_E$ greater than a few hundred 
thousand years, and provided that the emission processes last less than about $t_E$, the stationary limit considered here still gives accurate results. 

A second assumption adopted here is that of emitters that are distributed in the Milky Way independently of each other. Adding correlations
between the emitters would be functional to describe clustering effects arising, for example, by space-faring species colonizing nearby
planetary systems, as in the directed panspermia scenario \citep{Ginsburg2021}. In part, clustering can be mimicked by adopting \textit{ad hoc} functional forms of
the emitter PDF, as done in Section \ref{sec:model} where we used a more or less localized Gaussian for PDF. We note however that, 
given enough time, a possible outcome of directed panspermia is the colonization of the entire galaxy \citep{Carroll2019}. In this case, the emitters would be 
uniformly distributed over the Milky Way, as considered in this paper.    
  
In conclusion, we do not know whether the premise laid out in this paper (i.e., that technoemission have not crossed Earth since more than 60 years) 
is true or not, but it is certainly an hypothesis that needs to be considered,
especially after decades of fruitless searches and only two years before the Breakthrough Listen project is completed. 
This rises the question of whether SETI science should focus more on commensal investigations, i.e., searching for technosignals from data collected 
by telescopes performing other observational activities, rather than investing telescope time in active SETI searches.

\begin{acknowledgments}
The author wishes to thank A. Balbi, P. De Los Rios, J. Kuennen, M. Lingam and G. W. Marcy for advice and comments on early drafts.
\end{acknowledgments}

\appendix 
\section{Derivation of the likelihood function}
\label{AppA}

Our model considers a collection of statistically independent spherical shells, each representing a region of space filled by isotropic
electromagnetic radiations emitted from a random position in the galaxy, and a test sphere of diameter $\delta$ and center
at Earth's position $\mathbf{r}_o$. The spherical shells can overlap with each other and with the test sphere, so that the 
probability that $k$ shells overlap the test sphere follows a Poisson distribution: $\eta(\delta)^k e^{-\eta(\delta)}/k!$, where
$\eta(\delta)$ is the average number of overlaps.  Setting $k=0$ yields the probability that none of the spherical shells 
overlap the test sphere: $P(\delta)=e^{-\eta(\delta)}$.

To calculate $\eta(\delta)$ we consider the probability of a single shell overlapping the test sphere:
\begin{equation}
\label{pshell}
p(\delta; t, L)=\int\!\!d\mathbf{r}\,\rho_E(\mathbf{r})\theta(ct+\delta/2-d)\theta(d -ct+cL+\delta/2),
\end{equation}
where $d=\vert\mathbf{r}-\mathbf{r}_o\vert$ is the distance of an emitter from the Earth, $\theta(x)=1$ if $x\geq 0$ and $\theta(x)=0$ if $x<0$ is the 
unit step function, $\rho_E(\mathbf{r})$ is the probability density of an emitter being located in $\mathbf{r}$, $ct$ is the outer 
radius of the spherical shell and $cL$ its thickness, where $t\geq 0$ is the elapsed time since the emission started and $c$ is the 
speed of light. In the case of multiple shells that are generate with rate $\Gamma(t)$, the average number of overlaps is obtained 
by marginalizing \eqref{pshell} over $t$ and $L$. Exchanging the order of integration and noting that $\Gamma(t)=0$ for $t<0$ and 
that $\delta=c\tau$ we find:
\begin{eqnarray}
\label{etaapp}
\eta(c\tau)&=&\int\!dL \,\rho_L(L)\int\!dt\,\Gamma(t)p(c\tau; t, L)\nonumber \\
&=&\!\int\!\!dL \,\rho_L(L)\!\!\int\!\!d\mathbf{r}\rho_E(\mathbf{r})\bigg[\theta\!\left(\frac{d}{c}-\frac{\tau}{2}\right)
\!\int_{\frac{d}{c}-\frac{\tau}{2}}^{\frac{d}{c}+\frac{\tau}{2}+L}\!\!\!\!dt\,\Gamma(t)\nonumber \\
&&+\,\theta\!\left(\frac{\tau}{2}-\frac{d}{c}\right)\!\int_0^{\frac{d}{c}+\frac{\tau}{2}+L}\!\!\!\!dt\,\Gamma(t)\bigg].
\end{eqnarray}

\subsection{Stationary limit}
\label{AppA1}
Under the assumption that $\Gamma(t)$ does not change appreciably within the limits of integration over $t$ in Equation \eqref{etaapp},
we neglect the time dependence of the emission birthrate and set $\Gamma(t)= \Gamma$.
Performing the integration over $t$ and $L$ then yields:
\begin{equation}
\label{iso1}
\eta(c\tau)= \Gamma[\tau+\bar{L}-K(\tau)],
\end{equation}
where $\bar{L}=\int\!dL\,\rho_L(L)L$ is the average longevity of the emissions and
\begin{equation}
\label{iso2}
K(\tau)=\frac{1}{2}\int\!d\mathbf{r}\rho_E(\mathbf{r})\theta(\tau-2d/c)(\tau-2d/c).
\end{equation}
Equations \eqref{iso1} and \eqref{iso2} yield Equation~\eqref{etaa} of the main text. Finally, the conditional probability that the 
time between overlaps is greater than $\tau$, given $\Gamma$, reads:
\begin{equation}
\label{prob3}
P(\tau\vert \Gamma)=P(c\tau)/P(0)=e^{-\Gamma [\tau-K(\tau)]}.
\end{equation}

\subsection{First order corrections in $t$}
\label{AppA2}
To estimate the importance of the time dependence of the emission rate, we Taylor expand $\Gamma(t)$ up to the first order in $t$ and write
$\Gamma(t)\simeq \Gamma(1+t/t_E)$, where $t_E$ is some characteristic timescale. The time integration in Equation \eqref{etaapp} can still be
performed analytically, yielding for $P(\tau\vert \Gamma)$:
\begin{equation}
\label{probt1}
P(\tau\vert \Gamma)=e^{-\Gamma [\tau(1+(\bar{L}+2\bar{d}/c)/2t_E)-K(\tau)+K_1(\tau)/4t_E]},
\end{equation}
where $\bar{d}=\int\!d\mathbf{r}\rho_E(\mathbf{r})\vert \mathbf{r}-\mathbf{r}_o\vert$ is the mean Earth-emitter distance and
\begin{equation}
\label{probt2}
K_1(\tau)=\frac{1}{2}\int\!d\mathbf{r}\rho_E(\mathbf{r})\theta(\tau-2d/c)(\tau-2d/c)^2.
\end{equation}
The stationary limit of Equation \eqref{prob3} is recovered by setting $t_E\rightarrow\infty$. For $\tau$ smaller than $t_E$, the main contribution
of a non-stationary $\Gamma(t)$ comes from the factor $(\bar{L}+2\bar{d}/c)/2t_E$ in Equation \eqref{probt1}. Although this correction introduces 
an explicit dependence on $\bar{L}$ (absent in the stationary limit) it is negligible small as long as $t_E\gg\bar{L}$ and $\bar{d}/c$.

\subsection{Models of the emitter distribution $\rho_E(\mathbf{r})$}
\label{AppA3}
In the main text, we show results obtained by using two functional forms of $\rho_E(\mathbf{r})$. The first one adopts an axisymmetric distribution 
of the emitters of the form:
\begin{equation}
\label{GHZ}
\rho_E(\mathbf{r})=\lambda (r/r_s)^\beta\exp(-r/r_s)\exp(-\vert z\vert/z_s),
\end{equation}
where $r$ is the radial distance from the galactic center, $z$ is the height from the galactic plane, and $\lambda$ is a normalization factor. By setting
$\beta=0$, $r_s=8.15$ kly, and $z_s=0.52$ kly, Equation \eqref{GHZ} reproduces the distribution of stars in the thin disk of the Milky Way, whereas
for $\beta=7$ and $r_s=3.26$ kly it replicates the main features of the annular galactic habitable zone of \citet{Lineweaver2004}. 
An approximate but sufficiently accurate expression for $P(\tau\vert\Gamma)$ can be derived by substituting 
in Equations \eqref{iso2} the Earth-emitter distance $d=\vert \mathbf{r}-\mathbf{r}_o\vert$  for its mean $\bar{d}$, which gives Equation \eqref{etab}
of the main text.

In the second model, we consider a Gaussian function centered on $\mathbf{r}_E$ and with standard deviation $\sigma$: $\rho_E(\mathbf{r})=\exp(-\vert\mathbf{r}-\mathbf{r}_E\vert^2/2\sigma^2)/(2\pi)^{3/2}\sigma^3$. In this case, Equation \eqref{etab} (with $\bar{d}=\vert\mathbf{r}_E-\mathbf{r}_o\vert$) becomes
increasingly accurate as $\sigma/\bar{d}\rightarrow 0$.

\section{Anisotropic emissions}
\label{AppB}

We model a directional anisotropic technoemission by a conical beam of aperture $\alpha$ and
axis oriented along the direction of the unit vector $\mathbf{n}$. As done for the isotropic case, we take a test sphere of radius $\delta/2$ centered at Earth
and consider the probability that the beamed emission overlaps the test sphere. For $\mathbf{n}$ averaged uniformly over the unit sphere, this is given by:
\begin{equation}
\label{ani1}
p(\delta; t, L,\alpha)=\Omega(\alpha)p(\delta; t, L),
\end{equation}
where $\Omega(\alpha)=[1-\cos(\alpha/2)]/2$ is the fractional solid angle subtended by the beam and $p(\delta; t, L)$ is the overlap 
probability given in Equation \eqref{pshell}. 

Next, we denote with $\Gamma_\textrm{iso}$ and $\Gamma_\textrm{ani}$ the rate of isotropic and anisotropic technoemissions, respectively, so that 
using Equation \eqref{etaapp} the average number of emissions overlapping the test sphere of diameter $\delta=c\tau$ reduces to:
\begin{eqnarray} 
\label{ani2}
\eta(c\tau)&=&\Gamma_\textrm{iso}\!\int\!\!dL \,\rho^\textrm{iso}_L(L)\int\!dt\,p(c\tau; t, L)\nonumber \\
&+&\Gamma_\textrm{ani}\Omega(\alpha)\!\int\!\!dL \,\rho^\textrm{ani}_L(L)\!\int\!dt\,p(c\tau; t, L),
\end{eqnarray}
where $\rho^\textrm{iso}_L(L)$ and $\rho^\textrm{ani}_L(L)$ are the longevity PDFs assigned to the isotropic and anisotropic emissions, respectively.
The integration over $t$ and $L$ yields:
\begin{eqnarray}
\label{ani3}
\eta(c\tau)&=&\Gamma_\textrm{iso}\bar{L}_\textrm{iso}+\Gamma_\textrm{ani}\Omega(\alpha)\bar{L}_\textrm{ani}\nonumber \\
&&+[\Gamma_\textrm{iso}+\Omega(\alpha)\Gamma_\textrm{ani}][\tau+K(\tau)],
\end{eqnarray}
where $\bar{L}_i=\int\!dL\,\rho_L^i(L)L$ ($i=$ iso, ani) and $K(\tau)$ is defined in Equation \eqref{iso2}. Finally, setting $\Gamma=\Gamma_\textrm{iso}
+\Gamma_\textrm{ani}$, $q=\Gamma_\textrm{ani}/\Gamma$, and $\tau=\delta/c$, we obtain:
\begin{equation}
\label{ani4}
P(\tau\vert\Gamma)=e^{-\Gamma^*[\tau-K(\tau)]},
\end{equation}
where $\Gamma^*=\Gamma[q+(1-q)\Omega(\alpha)q]$. 

\bibliographystyle{aasjournal}

\begin{thebibliography}{99}

\bibitem[Balbi \& \'Cirkovi\'c(2021)]{Balbi2021}
Balbi, A., \& \'Cirkovi\'c, M. M. 2021, AJ, 161, 222

\bibitem[Balbi \& Grimaldi(2022)]{Balbi2022}
Balbi, A., \& Grimaldi, C. 2022, in Technosignatures for Detecting Intelligent Life in Our Universe: A Research Companion, 
ed. A. Berea (Hoboken, Wiley-Scrivener), 127

\bibitem[Carroll-Nellenback et al.(2019)]{Carroll2019}
Carroll-Nellenback J., Frank A., Wright J., \& Scharf C. 2019, AJ, 158, 117

\bibitem[\'Cirkovi\'c(2004)]{Cirkovic2004}
\'Cirkovi\'c, M. M. 2004, AsBio, 4, 225

\bibitem[Drake(1961)]{Drake1961}
Drake, F. D. 1961, Phys. Today, 14, 40

\bibitem[Dyson(1960)]{Dyson1960}
Dyson, F. J. 1960, Science, 131, 1667

\bibitem[Enriquez et al.(2017)]{Enriquez2017}
Enriquez J. E., et al. 2017, ApJ, 849, 104

\bibitem[Forgan(2019)]{Forgan2019}
Forgan, D. H. 2019, Solving Fermi's Paradox, Cambridge University Press, Cambridge, UK

\bibitem[Gajjar et al.(2021)]{Gajjar2021}
Gajjar, V., et al. 2021, AJ, 162, 33

\bibitem[Gajjar et al.(2022)]{Gajjar2022}
Gajjar, V., et al. 2022, ApJ, 932, 81.

\bibitem[Ginsburg \& Lingam(2021)]{Ginsburg2021}
Ginsburg I., Lingam M., 2021, Res. Notes AAS, 5, 154

\bibitem[Gott(1993)]{Gott1993}
Gott, J. R. III 1993, Nature, 363, 315

\bibitem[Grimaldi \& Marcy(2018)]{Grimaldi2018}
Grimaldi, C., \& Marcy, G. W. 2018, PNAS, 115, E9755

\bibitem[Grimaldi(2021)]{Grimaldi2021}
Grimaldi, C. 2021, MNRAS, 500, 2278

\bibitem[Lares, Funes \& Gramajo(2020)]{Lares2020}
Lares, M., Funes, J. G., \& Gramajo, L. 2020, IJAsB, 19, 393

\bibitem[Lin et al.(2014)]{Lin2014}
Lin, H., Abad, G. G., \& Loeb, A. 2014,. ApJL., 792, L7

\bibitem[Lingam \& Loeb(2021)]{Lingam2021}
Lingam, M., \& Loeb, A. 2021, Life in the Cosmos: From Biosignatures to
Technosignatures (Cambridge, MA: Harvard Univ. Press)

\bibitem[Lineweaver et al.(2004)]{Lineweaver2004}
Lineweaver, C., Fenner, Y., \& Gibson, B. 2004, Science, 303, 59

\bibitem[Kipping, Frank \& Scharf(2020)]{Kipping2020}
Kipping, D., Frank, A., \& Scharf, C. 2020, IJAsB, 19, 430

\bibitem[Kopparapu et al.(2021)]{Kopparapu2021}
Kopparapu, R., Arney, G., Haqq-Misra, J., Lustig-Yaeger, J., \& Villanueva, G. 2021, AJ, 908, 164

\bibitem[Price et al.(2020)]{Price2020}
Price D. C., et al. 2020, AJ, 159, 86

\bibitem[Rozwadowska, Vissani \& Cappellaro (2021)]{Rozwadowska2021}
Rozwadowska, K., Vissani, F., \& Cappellaro, E. 2021, New Astronomy, 83, 101498

\bibitem[Sellers et al.(2022)]{Sellers2022}
Sellers, L., Bobrick, A., Martire, G., Andrews, M., \& Paulini, M. 2022, arXiv: 2212.02065, MNRAS submitted

\bibitem[Shklovskii \& Sagan(1965)]{Shklovskii1965}
Shklovskii, I., \& Sagan, C.1966,Intelligent Life in the Universe (SanFrancisco,
CA: Holden-Day)

\bibitem[Smith(2009)]{Smith2009}
Smith, R. D. 2009, IJAsB, 8, 101

\bibitem[Spiegel \& Turner(2012)]{Spiegel2012}
Spiegel, D. S., \& Turner, E. L. 2012, PNAS 109, 395

\bibitem[Suazo et al.(2022)]{Suazo2022}
Suazo, M., et al. 2022, MNRAS, 512, 2988

\bibitem[Tarter et al.(2010)]{Tarter2010}
Tarter, J. C., Agrawal, A., Ackermann, R., Backus, P., Blair, S. K. M., Bradford, T., Harp, G. R., Jordan, J., Kilsdonk, T., Smolek, K. E., Richards, J., Ross, J., Shostak, G. S.,  Vakoch, D. 2010, Proc. SPIE, 7819, 781902

\bibitem[Tellis \& Marcy(2015)]{Tellis2015}
Tellis, N. K., \& Marcy, G. W. 2015, PASP, 127, 540

\bibitem[Tipler(1980)]{Tipler1980}
Tipler, F. J. 1980, Q. Jl. R. Astr. Soc., 21, 267

\bibitem[Torquato(2002)]{Torquato2002}
Torquato, S. 2002, Random Heterogeneous Materials: Microstructure and Macroscopic Properties (Springer, New York, 2002).

\bibitem[Townes(1983)]{Townes1983}
Townes, C. H. 1983, PNAS, 80, 1147

\bibitem[Ward \& Brownlee(2000)]{Ward2000}
Ward, P., \& Brownlee, D. 2000, Rare Earth: Why Complex Life is Uncommon in the Universe. Copernicus Book, (Springer, New York)

\bibitem[Wlodarczyk-Sroka et al. (2020)]{Wlodarczyk2020}
Wlodarczyk-Sroka, B. S., Garrett, M. A., \& Siemion, A. P. V., 2020, MNRAS, 498, 5720

\bibitem[Worden et al. (2017)]{Worden2017}
Worden S. P., et al. 2017, Acta Astronaut., 139, 98

\bibitem[Wright et al.(2018)]{Wright2018}
Wright, J. T., Kanodia, S., \& Lubar, E. 2018, AJ, 156, 260

\bibitem[Wright et al. (2022)]{Wright2022}
Wright, J. T., Haqq-Misra, J., Frank, A., Kopparapu, R., Lingam, M., \& Sheikh, S. Z. 2021, ApJL, 927, L30

\end{thebibliography}

\end{document}